# Observational consequences of fine structure line optical depths on infrared spectral diagnostics

Nicholas Abel, Adam Bryant, Prabodh Dhakal, Ashley Gale, Alva Gibson, William Goddard, Chad Howard, Ameya Kolarkar, Pey Lian Lim, Gargi Shaw, & Gary Ferland

Department of Physics and Astronomy, University of Kentucky, Lexington, KY 40506

npabel2@uky.edu , albrya0@hotmail.com , prabodhdhakal@hotmail.com , aagale2@uky.edu , apgibs0@pop.uky.edu , goddard@pa.uky.edu , chad_howard77@hotmail.com , askola2@uky.edu , klangjaya@hotmail.com , gshaw@pa.uky.edu , gary@pa.uky.edu

## Abstract

It has long been known that infrared fine structure lines of abundant ions, like the [O III] 88 μ line, can become optically thick in H II regions under certain high luminosity conditions. This could mitigate their potential as diagnostic tools, especially if the source is too dusty for optical spectroscopy to otherwise determine the system's parameters. We examined a series of photoionization calculations which were designed to push the nebulae into the limit where many IR lines should be quite optically thick. We find that radiative transfer effects do not significantly change the observed emission line spectrum. This is due to a combination of grain absorption of the hydrogen ionizing continuum and the fact that the correction for stimulated emission in these lines is large. Given these results, and the likelihood that real objects have non-thermal line broadening, it seems unlikely that line optical depth presents a problem in using these lines as diagnostics of the physical conditions or chemical composition.

# 1 Introduction

The infrared spectral region makes it possible to observe objects that may be heavily shrouded in dust, and to see familiar objects in new ways. For instance, the ultra luminous infrared galaxies (ULIRG) are among the most luminous objects in the universe (Sanders & Mirabel 1996), but are difficult to study using conventional optical emission lines methods (c.f., Osterbrock 1989) due to their high obscuration. Infrared spectroscopy must lead the way in understanding these and related phenomena. IR line spectroscopy provides new insights to such familiar objects as AGN (Sturm et al. 2002). In some objects little associated visible emission will be detectable and an analysis must rely solely on IR lines.

Rubin (1968) and Simpson (1975) showed that infrared forbidden lines can become optically thick under some conditions. This introduces an uncertainty that diminishes their usefulness for observational analysis. When transitions become optically thick line photons scatter before escape, and it is possible that ratios of forbidden lines will no longer indicate density and temperature, or reflect the chemical composition of the emitter. Under what conditions do forbidden lines become optically thick, and what impact does this have on the diagnostic power of the lines?

In this paper we explore a range of model H II regions, extending across plausible parameters, to determine when the line optical depths become large. We also show how this will affect conventional IR-line forbidden line diagnostics. This follows in the footsteps of previous investigations of the infrared forbidden lines done by Spinoglio & Malkan (1992) and Voit (1992).

## 2 Photoionization calculations

The purpose of this paper is to examine what happens to the infrared forbidden line spectrum where the lines are expected to become optically thick. We focus on lines that form within the $^3$P ground term of astrophysically abundant ions, since these are among the strongest lines and are commonly used density indicators. We first estimate the parameters where the lines will become optically thick and then examine the consequences by running a grid of photoionization simulations.

### 2.1 Optical Depth of the [OIII] 88.4μ Line

We first develop an analytical expression for the expected optical depth of the $^3P_0 - ^3P_1$ [O III] transition. We concentrate on this transition since the oxygen abundance is large and the line is strong, and so it should suffer the largest effects.

The hydrogen ionization balance for a plane parallel slab is given by (Osterbrock 1989)

$$\phi \geq n_e n_p \alpha_B l \quad [\text{cm}^{-2}\ \text{s}^{-1}] \tag{1.}$$

where $\phi$ is the flux of hydrogen-ionizing photons striking the cloud, $l$ is the Strömgren length of the H$^+$ zone, $n_e$ and $n_p$ are the electron and proton densities, and $\alpha_B$ is the case B recombination coefficient to all levels but the first. This inequality holds if dust is present and removes some of the ionizing radiation (see Bottorff et al. 1998). We can define an ionization parameter $U$ as,

$$U \equiv \frac{\phi}{n_H c} \quad . \tag{2.}$$

Then the ionization balance equation can be rewritten as

$$N_H \leq cU/\alpha_B \quad [\text{cm}^{-2}] \tag{3.}$$

where $N_H = l\, n_H$ is the hydrogen column density.

In general the optical depth $\tau$ of a line is:

$$\tau = a_\nu \left( N_l - N_u \frac{g_l}{g_u} \right) \tag{4.}$$

where $a_\nu$ is the line center absorption cross section, $n_l$ and $n_u$ are the populations of the lower and upper levels, and $g_l/g_u$ are the statistical weights of the lower and upper levels. If stimulated emission is neglected then the term in parentheses becomes simply $n_l$, which is equal to $n(O^{++})$ in this case. Then the optical depth is:

$$\tau \approx a_\nu N(O^{++}) = a_\nu N_H \left( \frac{n(O^{++})}{n(O)} \right) \left( \frac{n(O)}{n(H)} \right) \tag{5.}$$

where $n(O)/n(H)$ is the oxygen to hydrogen abundance ratio and $n(O^{++})/n(O)$ is the fraction of oxygen that is doubly ionized, averaged over the column.

The line center absorption coefficient is related to the oscillator strength by:

$$a_\nu = \frac{\sqrt{\pi} e^2 \lambda f_{ij}}{m_e c\, \Delta u_D} \; [\mathrm{cm^2}] \tag{6.}$$

where $\Delta u_D$ is the Doppler linewidth in velocity units.

Emission lines in H II regions often have widths that correspond to supersonic motions. This non-thermal line broadening can amount to several hundred km s[-1] in extragalactic H II regions (Melnick, Tenorio-Tagle, & Terlevich 1999). The nature of this turbulence and the physical scale this motion occurs on are not known. The velocity width that enters into equation 6. is given by:

$$\Delta u_D = \sqrt{\Delta u_{th}^2 + \Delta u_{turb}^2} \tag{7.}$$

with $\Delta u_{th} = \sqrt{2kT/m}$ is the line width due to thermal motions and $\Delta u_{turb}$ is the *micro*-turbulent line width. *Macro*-turbulence, due to bulk motions of entire clouds, does not add a turbulent term to equation 6, but *micro*-turbulence, in which gas motions occur over a scale of order the photon mean free path for scattering, does.

Here we assume that only thermal linewidths contribute to line broadening, to obtain the largest line absorption cross section and so maximize the effects of line optical depths. We carry through the ratio $\Delta u_{th}/\Delta u_D$, the ratio of the thermal to total doppler line width, as a reminder. Our assumption above makes this ratio equal to one, but this ratio would be $10^{-1}$ to $10^{-1.5}$ if the line widths observed in the extragalactic H II regions were due to microturbulence.

Substituting equation 3 into equation 5 gives the optical depth:

$$\tau \leq \left(\frac{\Delta u_{th}}{\Delta u_D}\right)\left[\frac{n(O)}{n(H)}\right]\left[\frac{n(O^{++})}{n(O)}\right]\frac{a_v cU}{\alpha_B} . \qquad (8.)$$

Substituting the atomic data for [OIII] 88.4μ, (Osterbrock 1989) assuming an oxygen abundance of $n(O)/n(H) = 3.2(-4)$, (Cowie and Sanaglia 1986; Savage and Sembach 1996) and assuming a temperature of $10^4$ K we find

$$\begin{aligned}\tau &\leq 4.5\times 10^5 \left[\frac{n(O)}{n(H)}\right]\left(\frac{\Delta u_{th}}{\Delta u_D}\right)\left[\frac{n(O^{++})}{n(O)}\right]U \\ &\leq 1.44\times 10^2 \left[\frac{n(O)/n(H)}{3.2\times 10^{-4}}\right]\left(\frac{\Delta u_{th}}{\Delta u_D}\right)\left[\frac{n(O^{++})}{n(O)}\right]U\end{aligned} . \qquad (9.)$$

For comparison, typical H II regions have $U \sim 10^{-2}$, so line optical depths cannot generally be ignored.

## 2.2 A series of model calculations

Next we will compute a series of blister-style model H II regions and examine the effects that line optical depth has on the predicted forbidden line diagnostics.

Version 96 of the spectral synthesis code Cloudy is used (Ferland 2002), and Bottorff et al. (1998) and Armour et al. (1999) give further details of our assumptions. This calculation includes the improved grain physics described by van Hoof et al. (2001). For comparison this produces less photoelectric heating than the previous simpler grain models. We use ISM abundances – a few relative to hydrogen are He/H = 0.098, C/H = 2.5(-4), N/H = 7.9(-5), O/H = 3.2(-4), Ne/H = 1.2(-4), and Ar/H = 2.8(-6). The geometry is given by Baldwin et al. (1991) – a plane parallel slab irradiated by a stellar continuum. The continuum source used is a 40,000 K blackbody. This was chosen for simplicity and because we do not expect our results to strongly depend on the continuum shape.

For simplicity, the hydrogen density was assumed to be constant. Models with $log(n(H))$= 2, 3, and 4 were computed. The line optical depth (equation 9) has no explicit dependence on density, but level populations do.

We examined a range of ionization parameters, or equivalently the ratio of the flux of H-ionizing photons $\phi$ to density. There is no lower limit to the value of $U$ that can occur, although very low ionization H II regions are not observed. Note that very few doubly ionized ions are present when $U < 10^{-2.5}$. There is an upper limit to U, however, because we assume that grains exist, which limits models to those in which the grain temperatures are below their sublimation points. Grains tend to equilibrate at a temperature a bit above the energy density temperature of the local radiation field, so the grain temperature is a function of $\phi$ rather than $U$. A typical grain sublimation temperature is on the order of $10^3$ K, therefore we can define a critical value, $\phi_{crit}$, such that at higher flux the grains will be too hot to survive. We find $\phi_{crit} = 2.46 \cdot 10^{19}$ [cm² s⁻¹] as the upper limit to the flux (and corresponding ionization parameter) that we consider.

Setting $\tau$ equal to 1 in equation 9 we can find a value $U_{crit}$ such that the [O III] lines are optically thick for values of $U \geq U_{crit}$:

$$U_{crit} \geq 6.9 \cdot 10^{-3} \left[ \frac{3.2 \cdot 10^{-4}}{n(O)/n(H)} \right] \left( \frac{\Delta u_D}{\Delta u_{th}} \right) \left[ \frac{n(O)}{n(O^{++})} \right]. \tag{10.}$$

All calculations were carried out over a range of U that exceeded this limit.

## 3 Results

Bottorff et al. (1998) show that there is a critical value of the ionization parameter above which grains, rather than hydrogen, will absorb ionizing radiation. Figure 1 shows the efficiency that the cloud converts ionizing radiation to ionized hydrogen for our calculations. When *U* is low then almost all of the radiation goes into ionizing hydrogen, while at large *U* most of the radiation is absorbed by dust and the cloud becomes "dust bounded". The hydrogen column density (equation 3) and associated optical depth (equation 9) are overestimates in this limit since the size of the H+ zone is overestimated. This will prevent the IR line optical depths from becoming as large as equation 9 suggests.

The calculations include a complete description of the grain physics, including charge, drift velocity, and temperature (van Hoof et al. 2001). Figure 2 shows the temperature of the average carbonaceous grain versus the flux of ionizing photons. Grains will sublimate when $\phi$ is greater than $10^{19}$ cm$^{-2}$ s$^{-1}$ due to their high temperature. This flux corresponds to an upper limit of $\log U \approx 5$ for $\log n(H) = 4$, the largest hydrogen density considered in our simulations, and $\log U \approx 7$ for $\log n(H) = 2$.

Figure 3 shows the optical depth of the [O III] lines versus *U* for three values of the hydrogen density. For smaller values of increasing *U* the optical depths increase as expected from equation 9, but for higher *U* the lines approach an asymptote that is not predicted by equation 9. Part of this is because, as mentioned above, the dust absorption that occurs at high values of *U* makes equation 9 an upper limit.

A further contributor is the fact that the correction for induced emission (equation 4) is not negligible for these lines at densities near or above their critical density. This means that the term $N_u(g_l/g_u)$ present in equation 4 cannot be neglected. The excitation temperature $T_{exc}$ is defined as:

$$n_u/n_l \equiv (g_l/g_u)\exp(-\chi/kT_{exc}), \tag{11}$$

where $\chi$ is the line excitation energy. When the density is at or above the critical density $T_{exc}$ will approach the gas kinetic temperature, which is given by the electron temperature $T_e$. We can rewrite equation 4 as

$$\tau = a_v N_l \left[1 - \exp(-\chi/kT_{exc})\right]. \tag{12}$$

Then we see that, since $T_e \sim 10^4$ K and $kT_{exc} \gg \chi$ we can expand equation 4 as

$$\tau \approx a_v N_l \chi / kT_{exc} \ll a_v N_l. \tag{13}$$

Hence the line optical depths are much smaller than expected.

Multiple scattering of marginally optically thick transitions will have a similar effect at lower densities – the population of the upper level relative to the lower level increases, making the correction for stimulated emission large. All of these processes prevent the line optical depth from ever becoming large.

We made several plots showing the ratio of the stronger to weaker line for the five $^3P$ ions with greatest abundance (Figure 4 a-e). This ratio can be used as a density indicator for some conditions (Osterbrock 1989). These plots show that these ratios remain valid density indicators, even for conditions where we would have expected that the lines would become optically thick.

The top two plots, the neon and argon ratios, show that the ratios of line intensities start to increase significantly for large values of *U*. These transitions have the highest critical densities of those shown, and so they are well below their critical densities, and so $T_{exc} \ll T_e$, for the parameters shown. This means that the correction for stimulated emission is small, and so optical depths do affect the ratios, but only at the very largest *U*. Objects with *U* this large would have remarkably high dust temperatures, close to their sublimation point.

## 4 Conclusions

The possibility that infrared fine structure lines could become optically thick mitigated their utility by introducing a basic uncertainty. We have presented calculations which span the full density and ionization range that occurs in H II regions. These included parameters where the IR lines should be very optically thick.

With few exceptions, optical depth effects are not as important as one might have thought. First, the Strömgren length does not increase monotonically with increasing *U* because of grain absorption of the incident continuum. This makes gas column densities and optical depths smaller than predicted by simple estimates. Second, the correction for stimulated emission is very large for infrared transitions at nebular temperatures, and this again makes the line opacity and optical depths much smaller. On top of this, real H II regions have non-thermal components to their line widths, which may further reduce the optical depth – our calculations assumed thermal broadening only.

For all of these reasons, the IR lines are largely optically thin, and should be a useful diagnostic of the physical conditions in most reasonable circumstances.

This work has been supported by NSF through grant AST-0071180 and by NASA with NAG5-12020 and NAG5-8212. We thank the referee for a careful review of the manuscript.

# 5 Figures

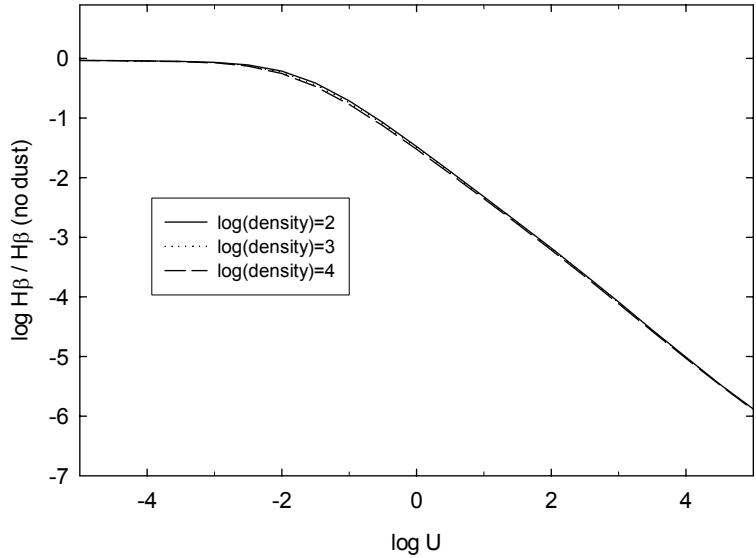

**Fig. 1**. The efficiency with which the cloud converts ionizing radiation into hydrogen recombination lines is shown versus ionization parameter. This shows the luminosity actually emitted by Hβ divided by the Hβ emission that would have occurred had no dust been present. For low ionization parameters almost all ionizing radiation is absorbed by hydrogen and produces hydrogen recombination lines. As $U$ increases a greater fraction of the incident starlight is absorbed by grains rather than hydrogen, the nebula becomes "dust bounded", and the cloud becomes predominantly an infrared emitter. Analytical calculations (Bottorff et. al. 1998) show that this occurs when $U \approx 10^{-2}$ for a galactic dust to gas ratio.

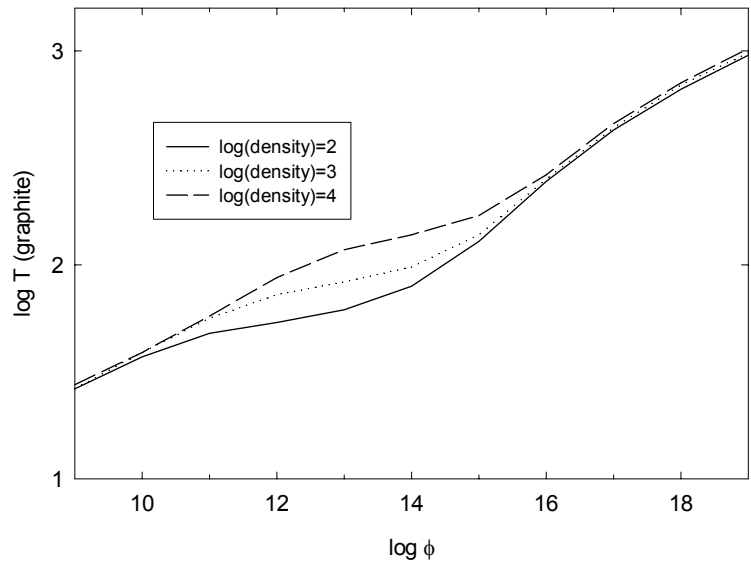

**Fig. 2.** The temperature of the carbonaceous dust component is plotted versus the flux of ionizing photons. Our calculations do not extend beyond the upper limit of $\phi_{crit} \approx 10^{19}$ cm² s⁻¹ that is set by the requirement that the grains survive. Dust temperature increases with increasing flux and the dust would sublimate when $\phi > \phi_{crit}$. The simple estimate of the energy density temperature given in the text generally reproduces this curve. Deviations are due to a combination of grain opacity effects and grain – gas collisions.

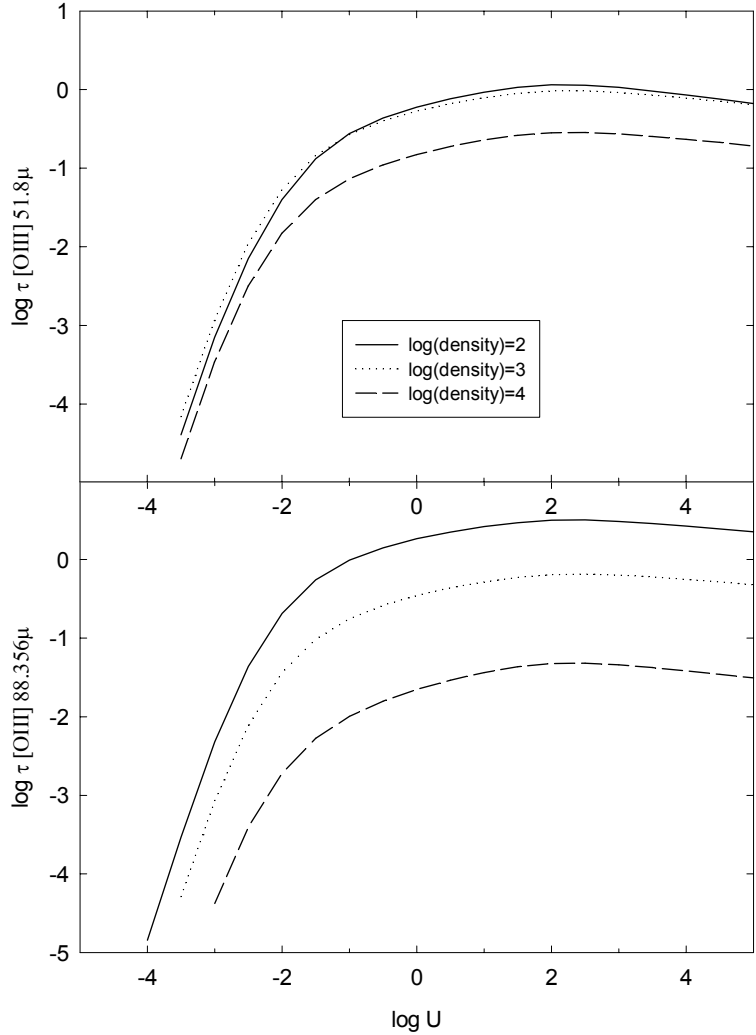

**Fig. 3a-b.** The optical depth for the [O III] $\lambda\lambda$ 51.8, 88.356µ lines is shown as a function of the ionization parameter. The simple estimate given in the text predicts that the optical depth should increase linearly with $U$ and reach unity at $U_{crit} \sim 10^{-2}$. Note that the optical depth actually approaches an asymptote at large values of $U$. This is because of grains absorbing the incident continuum and the very large correction to the line opacity due to simulated line emission.

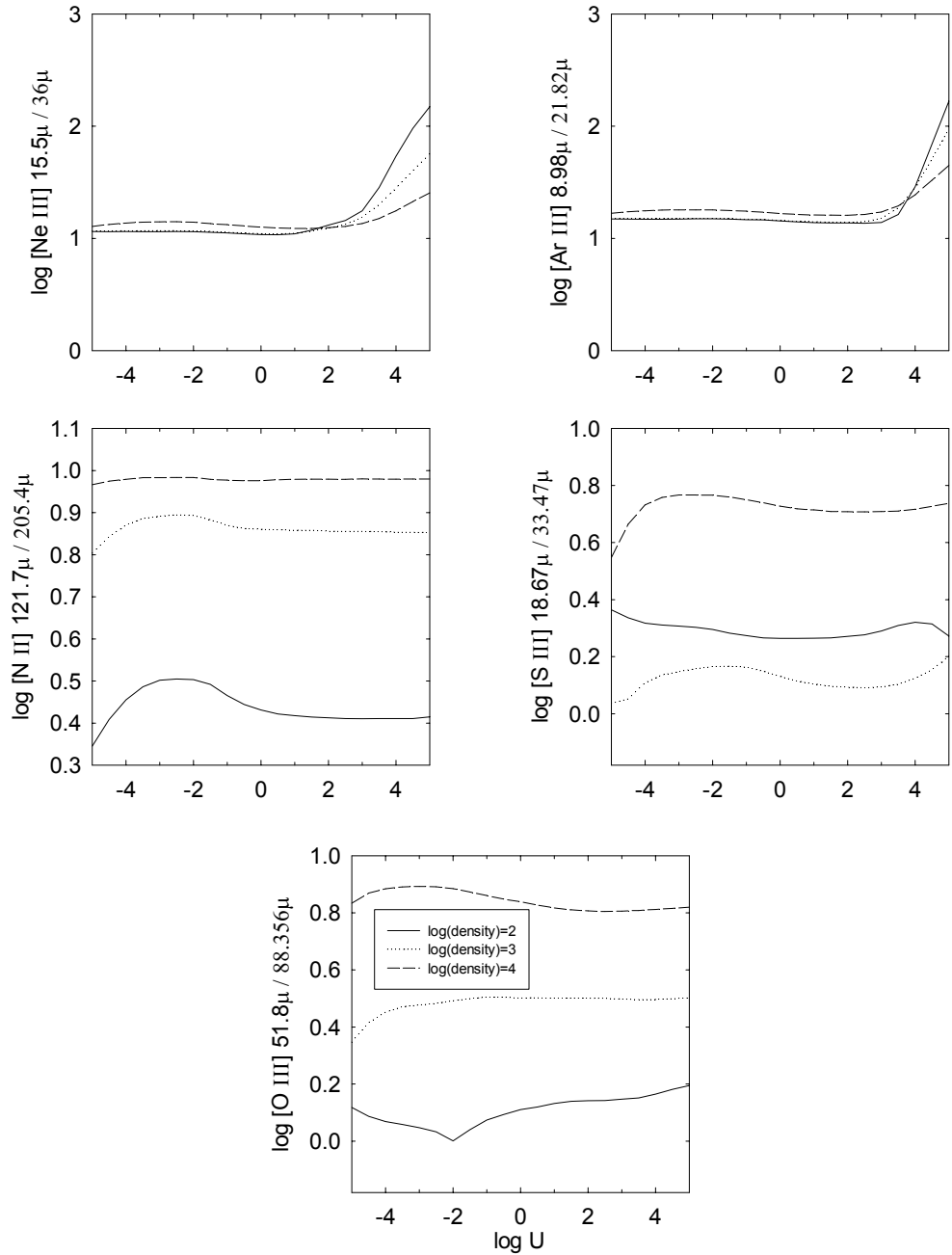

**Figs. 4a-e.** These figures show ratios of lines from the $^3P$ ground terms of five ions. They give the ratio of the intensity of the stronger line relative to the weaker line as a function of the ionization parameter. [Ne III] $\lambda\lambda$ 15.5, 36µ; [O III] $\lambda\lambda$ 51.8, 88.356µ; [N II] $\lambda\lambda$ 121.7, 205.4µ; **[**S III] $\lambda\lambda$ 18.67, 33.47µ; and [Ar III] $\lambda\lambda$ 8.98, 21.82μ are plotted. These ions are not the predominant stage of ionization for all values of $U$ – for instance oxygen is only predominantly doubly ionized when $\log(U) \geq -2.5$.